\documentclass[iop,numberedappendix]{emulateapj}
\usepackage{amsmath}
\usepackage{bm}
\usepackage{times}
\newcommand{\Fig}[1]{Figure~\ref{#1}}
\newcommand{\Eq}[1]{Equation~(\ref{#1})}
\newcommand{\EQ}{\begin{equation}}
\newcommand{\EN}{\end{equation}}
\newcommand{\vv}{\mbox{\boldmath $v$} {}}
\newcommand{\uu}{\mbox{\boldmath $u$} {}}

\newcommand{\Sec}[1]{Section~\ref{#1}}

\newcommand{\St}{\text{St}}
\newcommand{\DDD}{{\cal D} {}}

\graphicspath{{.}{./Figures/}}

\date{\today,~ $ $Revision: 1.20 $ $}
\begin{document}

\title{Partitioning tungsten between matrix precursors and chondrule precursors through relative settling}
\shorttitle{Partitioning tungsten}

\author{Alexander Hubbard\altaffilmark{1}}
\altaffiltext{1}{American Museum of Natural History, New York, NY, USA}
\email{{\tt ahubbard@amnh.org}}

\begin{abstract}
Recent studies of chondrites have found a tungsten isotopic anomaly between
chondrules and matrix. Given the refractory nature of tungsten, this implies that W was carried into
the solar nebula by at least two distinct families of pre-solar grains.  The observed chondrule/matrix
split requires that the distinct families were kept separate during the dust coagulation process,
and that the two families of grain interacted with the chondrule formation mechanism differently.
We take the co-existence of different families of solids in the same general orbital region
at the chondrule-precursor size as given, and explore the requirements for
them to have interacted with the chondrule formation process at significantly different rates. We show that this sorting of families of solids
into chondrule and matrix destined dust
had to have been at least as powerful a sorting mechanism as the relative settling of aerodynamically distinct grains at
at least two scale heights above the midplane. The requirement that the chondrule formation mechanism was
correlated in some fashion with a dust grain sorting mechanism argues strongly for spatially localized
chondrule formation mechanisms such as turbulent dissipation in non-thermally ionized disk surface layers,
and argues against volume filling mechanisms such as planetesimal bow shocks.
\end{abstract}



\section{Introduction}

Chondrites are a class of relatively unprocessed undifferentiated meteorites which inform us
about the components from which, and the environment in which, they were assembled.
They are named after chondrules, sub-mm inclusions which were melted at temperatures above $1700$\,K  \citep{1997AREPS..25...61H},
and which contribute a significant mass and volume fraction of their hosts. The other major
constituent of chondrites is fine-grained matrix, which was never melted and mostly remained cold \citep{2003GeCoA..67.4823H}.
Interestingly, the elemental composition of
matrix and chondrules are different, even within a given chondrite \citep{2016GeCoA.172..322E}. This is significant because 
particles the size of chondrules were presumably
agglomerations of sub-micron interstellar dust grains \citep{Draine03}. Any scatter in the composition of the sub-micron grains
should have been washed out once averaged by the billion. Even in the case of very rare elements,
which might have been carried by only a handful of initial grains, a handful of initial grains among billions could not
have determined whether their host agglomeration went on to be melted into a chondrule or not.

The chondrule heating process itself surely explains some of the moderately volatile abundance differences between chondrules and matrix
\citep{1991GeCoA..55..935H, Depletion}.
However, compositional differences have also been found for quite refractory species including
Ca, Al and Ti \citep{2016GeCoA.172..322E}. Further, \cite{2015LPI....46.2262B} found isotopic differences between highly refractory
tungsten found in Allende's chondrules compared to its matrix. Those differences are not due to hafnium decay but instead are well fit by
differing abundances of r- and s-process nucleosynthesis products.
It is difficult to imagine how thermal or chemical processing during chondrule
formation could have generated such isotopic heterogeneities in such a refractory species.
This suggests that the elemental and isotopic compositions of chondrule-precursors differed
from those of matrix-precursors, and that they differed for reasons beyond thermal or chemical processing.

This puzzle is exacerbated by the fact that  protoplanetary disks, of which the solar nebula
has no reason to have been an exceptional example, are expected to be vigorously turbulent and well mixed \citep{2015MNRAS.452.4054G}.
That means that any compositional gradients in the solar nebula had to have been shallow, and
it would have taken long time scales for the accretion flow through the disk to have brought in gas-dust mixtures with meaningfully
different abundances.
Chondrules are small and extremely well coupled to the gas \citep{2012Icar..220..162J,2015Icar..245...32H}.
Accordingly, even if there was a large scale isotopic gradient in the disk,
nearly all the chondrules formed at one location would have been carried away by turbulent diffusion or the accretion flow itself
by the time sufficient accretion had occurred for the local abundances to have changed.

It has also been found that different chondrite classes, with
very different chondrule to matrix ratios, and matrix and chondrules with significantly different abundances,
nonetheless have similar bulk abundances in a phenomenon known as complementarity \citep{2010E&PSL.294...85H}.
This makes
invoking large scale abundance gradients to explain the differences between chondrules and matrix abundances even less plausible:
if chondrules and matrix were formed from solids contained within
different volumes of gas, and had very different abundances, then how could they have been subsequently combined in the correct (but wildly varying)
ratios for their bulk abundances to approximate chondritic? Indeed, \cite{2015MNRAS.452.4054G} showed that complementarity cannot
then be maintained for prolonged time scales, and that matrix and chondrules must have been in some fashion co-genetic.

The implications of complementarity, and how strong the relationship between matrix and chondrules is, is not yet fully certain.
Bulk compositions of chondrites certainly vary at the ten percent level \citep{2010E&PSL.294...85H}. It is also clear that different chondrite
classes drew from separate reservoirs within the solar nebula; and it seems likely that
different chondrites within a given class did so as well \citep{2012M&PS...47.1176J}. 
Refractory titanium carries an isotopic signature which shows modest scatter between chondrite
classes \citep{2009Sci...324..374T}. Nonetheless, complementarity seems to be a particularly strong hypothesis for tungsten isotopes,
with Allende's bulk composition falling neatly in line with the rest of the inner solar system's \citep{2016LPI....47.1453B}.

\section{Motivation}

How could chondrules and matrix have ended up with differing isotopic signatures in highly refractory elements
as has been found experimentally for tungsten? The highly refractory nature of tungsten, and the nucleosynthetic nature of the W isotopic signature both imply that the signal
cannot be due to either thermal or chemical processing in the solar nebula, or due to hafnium decay.
This is particularly important because the chondrule formation process involves extreme temperatures and conditions \citep{1997AREPS..25...61H}.
Because it did not originate in the solar
nebula, it must have been carried into the solar nebula by pre-solar grains of differing origins. 
For us to see a pre-solar grain based isotopic signature between different chondritic meteorite components, the different,
submicron-scale pre-solar grains had to have experienced the collisional dust growth process differently. If they didn't, then
averaging over billions of pre-solar grains would have washed out any isotopic signal.

One way to achieve this would be if the grains
were found in different regions, or were injected into the disk at different times, when differing 
dust dynamics dominated. This would also allow an easy explanation of the different fates of the grains: chondrule formation temporal
or spatial intermittency, and matches well with the observations that chondules in different chondrite classes, and even within the same
chondrite class, were drawn from different reservoirs within the solar nebula \citep{2012M&PS...47.1176J}. Those reservoirs even
differed in refractory isotopes \citep[e.g.~titantium,][]{2009Sci...324..374T}.
However, the separation of tungsten isotopes between chondrules and matrix
seen Allende is extremely large compared to variations in the bulk tungsten isotope ratios of inner solar system bodies, including itself \citep{2016LPI....47.1453B}.
Thus, Allende is tungsten-isotope-complementary with the inner solar system.
Therefore, spatial or temporal differences in the pre-solar grain injection
violates complementarity: as found by \cite{2015MNRAS.452.4054G}, for matrix and chondrules to have had correlated composition, they
had to have been co-genetic.

Because of the complementarity constraints, the two dust grain categories with different W isotopic signatures had to have
spatially and temporally overlapped. Any other solution would need to explain why Allende has exactly the correct
chondrule volume fraction to match the inner solar system's tungsten isotope ratios, while chondrites as a class
have strongly varying chondrule volume fractions.
This requires two separate effects. Firstly, the W isotopes had to have been split between two categories of dust agglomerations. Henceforth we
will use ``dust'' to refer to solids in general, and ``dust grain'' for individual objects. Secondly,
those two categories had to have been destined to different, chondrule or matrix, fates.
We leave the details of partitioning elements and isotopes
between different families of dust grains
to a companion paper \citep[hereafter Paper I]{Fe-Col} and focus here on the different destinies of the families.

If chondrule heating events occurred sporadically but reasonably spatially evenly across broad reaches of
the solar nebula then they should have hit regions which selected for all types of dust compositions.
If, however, chondrule forming events were highly spatially localized, then differences in dust particle aerodynamics,
which would have allowed for spatial separation 
of agglomerations of differing compositions on small enough scales to not
disrupt complementarity \citep{2002ApJ...581.1344T}, would have also allowed for the creation of the W isotope signature.
As a hypothetical scenario with spatially localized chondrule melting we
focus on vertical settling, assuming that chondrule forming events occurred primarily high above the disk midplane: more
settled dust agglomerations would have been less likely to have been melted.
Many elements are partitioned between matrix and chondrules \citep{2016GeCoA.172..322E};
and the process we will develop would not only explain W isotopic signatures, but would drive variations in any abundances
which were correlated with the dust aerodynamical properties.  We focus on the tungsten isotopic signature only because
it is nearly invulnerable to thermal and chemical processing.

Tungsten is extremely rare \citep[about $10^{-7}$ by mass,][]{2003ApJ...591.1220L} so cannot have caused any difference in the dust fate
by itself. Instead, W must trace a more significant difference between the dust families. In Paper I we noted
that iron is also partitioned between chondrules and matrix.
It is easy to imagine that different tungsten isotopes, created in different nucleosynthetic environments,
would have been found in pre-solar grains of differing iron metal abundances.
This led us to  propose that ferromagnetic interactions between dust grains \citep{1967Natur.215.1449H} both
helped preserve primordial inhomogeneities, and also allowed the different dust families to grow to different sizes, with different
aerodynamics. 

Even though Fe is strongly partitioned between matrix and chondrules,
which inspired Paper I, that partitioning is certainly at least in part due to thermal and chemical processes,
e.g.~during the extreme conditions of chondrule formation itself \citep[e.g.][]{2005ASPC..341..407C}.
However, iron's ferromagnetism depends on many parameters, so while W isotopes might have been correlated
with ferromagnetically active iron, they need not have been correlated with bulk (metallic and non-metallic) iron. Further, because
the isotopic signature must have been pre-solar in origin, any other elemental or isotopic signatures implied by any model for the W isotope
partitioning would be nucleosynthetic, rather than chemical (i.e. sideophile) in origin unless the model also
invokes chemical processing.

While radial pressure perturbations, such as those associated with zonal flows \citep{2009ApJ...697.1269J},
can also sort dust agglomerations by size, settling is particularly attractive
because the vertical structure of disks has the same lifetime as the disk itself, avoiding requiring temporal correlations between
chondrule formation
and dust spatial sorting. Also, the solar nebula's surface layers are expected to have been
more magnetic active due to X-ray and FUV ionization \citep{1996ApJ...457..355G}, so magnetized turbulent dissipation models for chondrule
formation can be plausibly vertically localized.
As a final note, restricting chondrule formation to having
occurred at altitude would help explain why chondrules are so small \citep{2015Icar..245...32H}.
Interestingly, altitude restrictions would strongly constrain chondrule formation processes. For example, planetesimal
bow shocks \citep{1998M&PS...33...97H,2012ApJ...752...27M}
can only be produced by populations of highly eccentric planetesimals, whose orbits would presumably also often be highly inclined.
Accordingly, if it turns out that chondrule formation was indeed altitude restricted, we can rule out planetesimal bow shocks as the dominant
chondrule formation process.

\section{Dust vertical distribution}
\label{Sec_DVD}

If chondrule formation occurred only above a critical height $z_c$, then only fraction of the dust vertically mixed above that
height would have been melted to make chondrules. 
Here we calculate the vertical distribution of dust in a vertically isothermal disk in hydrostatic equilibrium following
\cite{2002ApJ...581.1344T}.
A particle moving through gas feels a drag acceleration:
\EQ
\left.\frac{\partial \vv}{\partial t}\right|_{\text{drag}} = - \frac{\uu - \vv}{\tau},
\EN
where $\vv$ is the particle's velocity, $\uu$ the gas velocity at the particle's position and $\tau$ is
a drag time scale.
Dust particles smaller than the local
gas mean-free-path (which includes chondrules and their precursors)
experience Epstein drag regime, with corresponding drag times:
\EQ
\tau_E =\sqrt{\frac{\pi}{8}}\frac{a \rho_s}{\rho_g v_{th}}, \label{taudef}
\EN
where $a$ are the particles' radii, $\rho_s$ their solid density, $\rho_g$ the local gas density, and
\EQ
v_{th}=\sqrt{\frac{k_B T}{m_m}}
\EN
the gas thermal speed and $m_m$ the gas molecular mass. Vertically isothermal disks in hydrostatic equilibrium have
\EQ
\rho_g (z) = \rho_{g0} e^{-z^2/2H^2} \label{rhog}
\EN
where $\rho_{g0}$ is the midplane gas density, $z$ is the height above the midplane,
$\Omega_K$ is the local Keplerian frequency and $H \equiv v_{th}/ \Omega_K$ the local
gas pressure height.

In the case of protoplanetary disks it is customary to define a Stokes number
\EQ
\St \equiv \tau \Omega_K, \label{St_def}
\EN
where $\Omega_K$ is the local Keplerian frequency. This Stokes number describes the degree to which
the dust particle can slip through the gas on orbital time scales, which are also expected to be turbulent time scales.
Combining Equations~(\ref{taudef}),(\ref{rhog}) and
(\ref{St_def}) we find
\EQ
\St = \St_0 e^{+z^2/2H^2} \label{EQ_St}
\EN
where
\EQ
\St_0 =\frac{\pi}{2} \frac{a \rho_s}{\Sigma_g}
\EN
is the Stokes number at the midplane.
For scale,
at $R=2.5$\,AU in an MMSN \citep{1981PThPS..70...35H}, an $a=0.25$\,mm, $\rho=3$\,g\,cm$^{-3}$ chondrule would have had
$\St_0 \lesssim 3 \times 10^{-4}$. Chondrule precursors, being porous, would have had significantly
lower midplane Stokes numbers. As we next develop, this value for $\St_0$ is low enough that we expect
strong vertical mixing in any scenario which includes sufficiently vigorous surface layer turbulence to
drive chondrule formation \citep{2007ApJ...670..805O}.

To lowest order in $z$, where $z$ is the height above the midplane, the vertical force
of gravity in a protoplanetary disk is
\EQ
g_z = - z \Omega_K^2.
\EN
The small value of $\St_0$ estimated above implies that
chondrule and chondrule-precursor sized agglomerations were well enough coupled to the gas that they can be assumed to have reached terminal velocity,
with
\EQ
v_z = -z \Omega_K^2 \tau= -\St \frac{\,z}{H}v_{th}, 
\EN
which drives a dust settling flux of
\EQ
F_S = v_z \rho  = - \St\frac{\,z}{ H} \rho\, v_{th}, \label{FFS}
\EN
where $\rho$ is the dust fluid density.

Protoplanetary disks are expected to be turbulent, with a corresponding turbulent diffusivity
for the dust $\DDD$ which we scale through a parameter $\beta$:
\EQ
\DDD \equiv \beta v_{th} H,
\EN
Note that our $\beta$ is approximately, but not identically,
the $\alpha$ of an $\alpha$-disk \citep{1973A&A....24..337S}, and the details of how turbulence drives or is driven by
angular momentum transport, and any anisotropy in the turbulence, could reasonably lead to $\alpha$ and $\beta$ differing
by factors of a few.
Turbulent diffusion drives a diffusive vertical dust flux
\EQ
F_D = - \rho_g \DDD \partial_z \left(\frac{\rho}{\rho_g}\right), \label{FFD}
\EN
where $\rho_g$ is the gas density. A vertical steady state is reached when the total vertical flux is zero:
\EQ
F = F_S +F_D =0. \label{FF}
\EN
Combining Equations~(\ref{EQ_St}), (\ref{FFS}), (\ref{FFD}) and (\ref{FF}),
we finally arrive at:
\EQ
\rho(z) = \rho(0) \exp \left[ -\frac{z^2}{2H^2} - \frac{\St_0}{\beta} \left( \exp\frac{z^2}{2H^2} -1\right)\right]. \label{rhoz}
\EN
The dust midplane density $\rho(0)$ can be found in terms of the dust surface density $\Sigma_d$ by using
\EQ
\Sigma_d \equiv \int_{-\infty}^{\infty} dz\, \rho(z). \label{Sigmad}
\EN

Note that in the derivation of \Eq{rhoz} we have assumed that $\beta$ is a function neither of $z$ nor of $\St$.
\Eq{rhoz} implies that the settling criteria is
\EQ
\frac{\St}{\beta} = \frac{\St_0}{\beta} \exp \frac{z^2}{2H^2}   \gtrsim 1,
\EN
We expect $\beta \sim \alpha \ll 1$, so dust should be extremely settled well below heights at which $\St \sim 1$. Accordingly,
only regions with $\St \ll 1$ contribute, and the assumption
that $\beta$ is not a function of $\St$ is well founded \citep{2007Icar..192..588Y}.
We are explicitly assuming that chondrule formation events are correlated with height, and the dissipation of magnetized turbulence
is an obvious possible explanation for that: poorly ionized midplanes are not expected to be strongly turbulent \citep{1996ApJ...457..355G}. In this scenario
$\beta$ would be expected to vary with altitude.
However, the value of $\beta$ does not matter in determining $\rho_d(z)$
as long as $\St/\beta \ll 1$.  Our approximation of $\beta$ being independent of $z$ therefore extends to 
\EQ
\frac{\St(z)}{\beta(z)} \ll 1 \label{deadness_limit}
\EN
anywhere in the vertical column below the chondrule formation height, i.e.~for all $|z|\ll z_c$.

\begin{figure}\begin{center}
\includegraphics[width=\columnwidth]{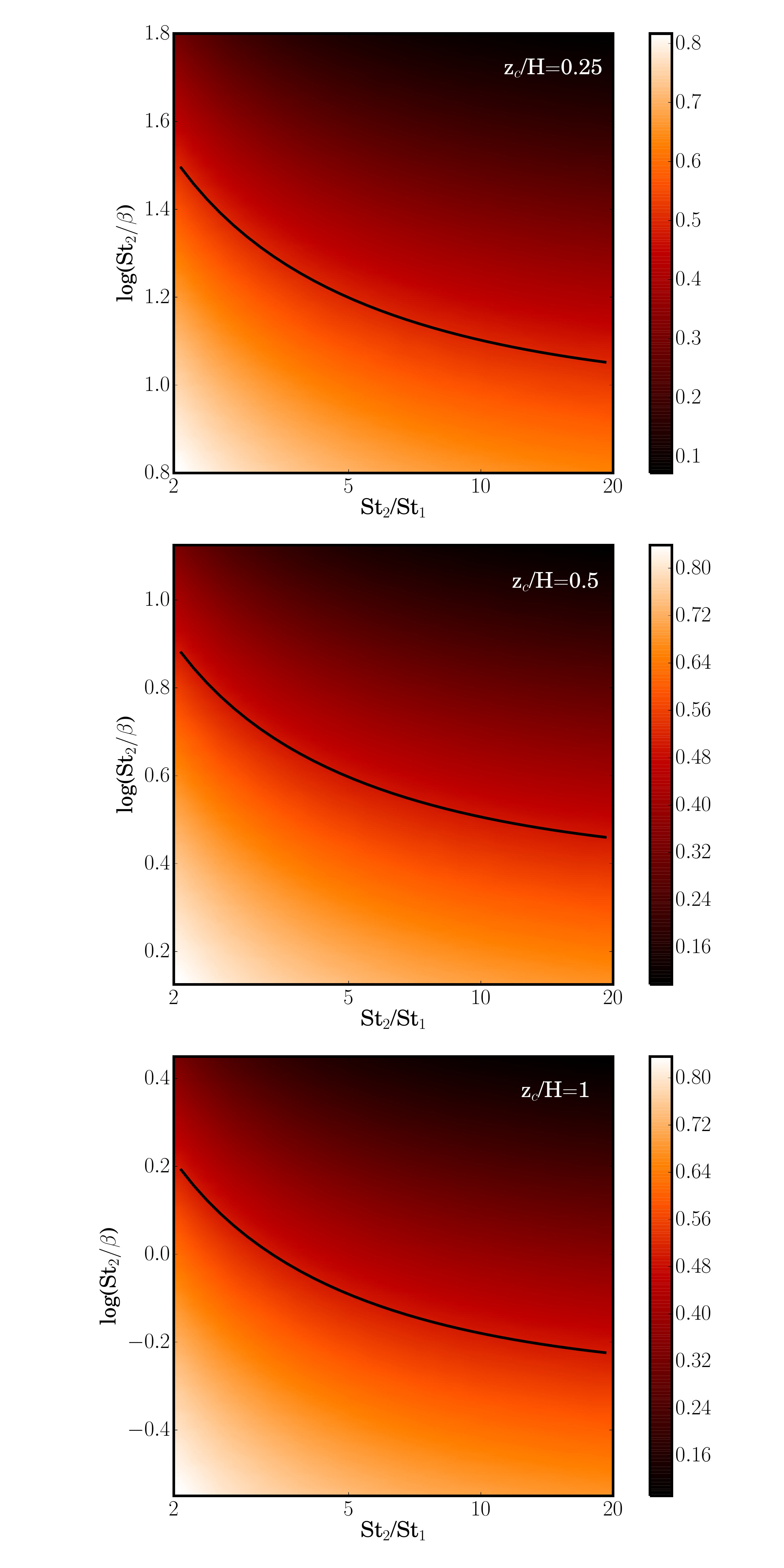}
\end{center}\caption{
Chondrule processing ratio $\xi$ (Equation~\ref{xi_def}) as a function of $\St_2/\St_1$ and $\St_2/\beta$
for $z_c/H =0.25, 0.5, 1$. Contour is $0.5$. Note that the x-axis is logarithmic.
\label{Scale_rat_large_fig} }
\end{figure}

If there was a sufficiently thick and dead midplane dead-zone,
then \Eq{deadness_limit} would not be satisfied,
and all the potential chondrule precursors would settle into a very thin layer at the midplane.
This is unlikely, as the low micro-physical hydrodynamical viscosity (as opposed to a turbulent effective viscosity) of 
protoplanetary disks means that any turbulent stirring in the surface layers would penetrate to the midplane 
with strength sufficient to mix $\St_0 \sim 10^{-4}$ particles \citep{2007ApJ...670..805O}.
As we will show in \Fig{Scale_rat_large_fig}, large
aerodynamical variations between dust families could still drive relative settling in the extremely low turbulence case;
but it is hard to see how any chondrule melting
processes active in such a laminar region of the disk, such as planetesimal bow shocks, would treat different altitudes differently.
Accordingly, we retain the assumption that \Eq{deadness_limit} is satisfied, which places qualitative lower bounds on $\alpha$ and $\beta$.
We emphasize that the strong height dependence of $\St$ (Equations~\ref{EQ_St}),
along with the small $\St$ of chondrules, and the even smaller $\St$ of their porous precursors,
nonetheless allows for even quite deadish dead zones.

\begin{figure}\begin{center}
\includegraphics[width=\columnwidth]{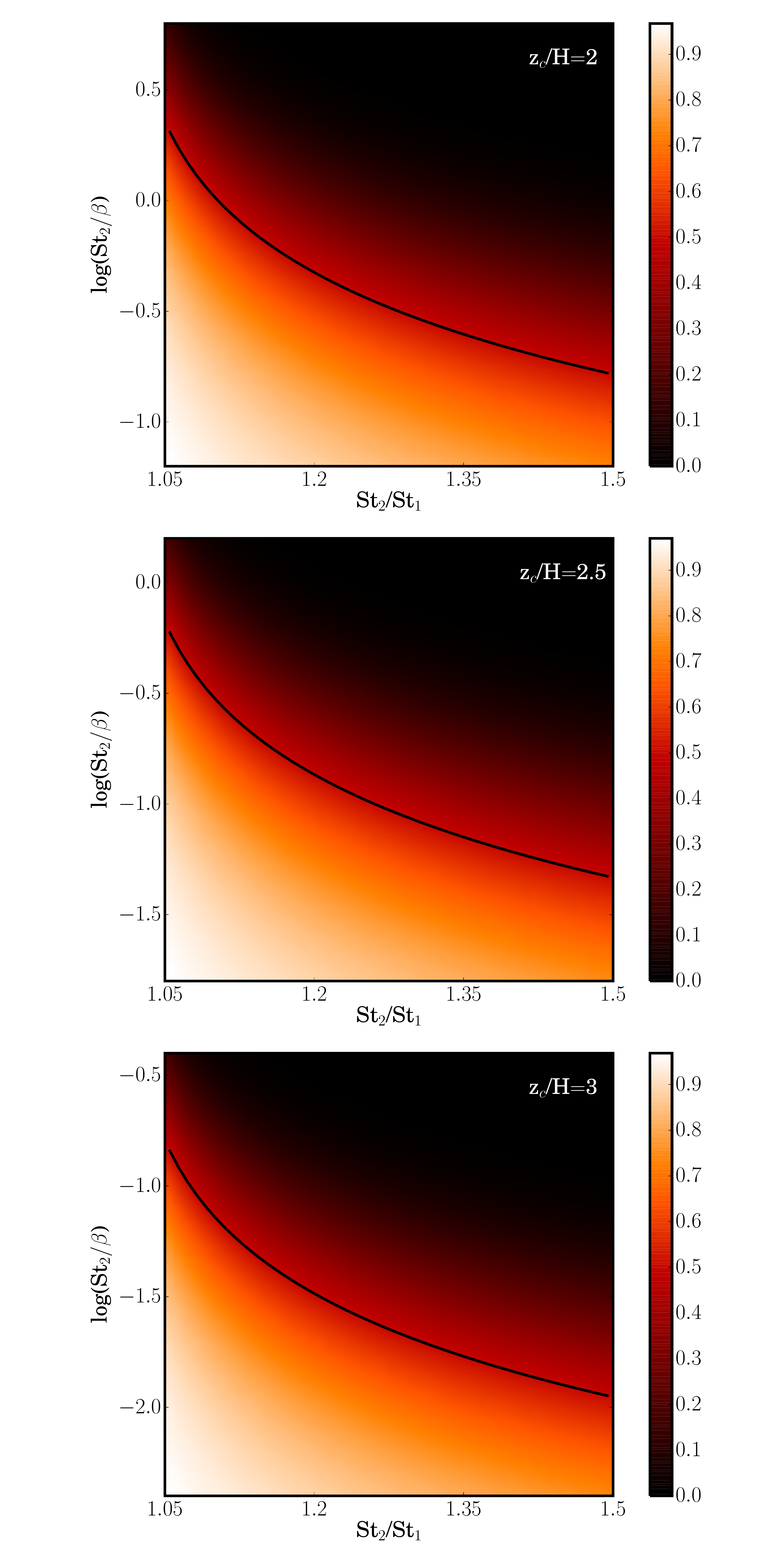}
\end{center}\caption{
Chondrule processing ratio $\xi$ (Equation~\ref{xi_def}) as a function of $\St_2/\St_1$ and $\St_2/\beta$
for $z_c/H =2, 2.5, 3$.  Contour is $0.5$. Note that the x-axis is linear.
\label{Scale_rat_small_fig} }
\end{figure}

\section{Chondrule melting fractions}

We assume the existence of two compositionally distinct dust agglomeration populations in the solar nebula whose aerodynamics were different, with
midplane Stokes numbers $\St_1 < \St_2$.
In future studies of chondrule and matrix abundances, the aerodynamical difference would be due to how the compositional variation affected
the coagulation processes, so the ratio $\St_2/\St_1$ would be a known constraint.
We also assume that chondrule melting events only occurred above a critical height $z_c$. We can use the vertical dust distributions calculated
above to determine the fraction of the the dust populations which were above $z_c$, and hence the different rates at which the two populations
were converted into chondrules.

From Equations~(\ref{rhoz}) and (\ref{Sigmad}) the fraction of dust population $i$ which is found in the chondrule forming region is
\EQ
f_i = \frac{2\int_{z_c}^{\infty} dz\,  \exp \left[ -\frac{z^2}{2H^2} - \frac{\St_i}{\beta} \left( \exp\frac{z^2}{2H^2} -1\right)\right]}
{ \int_{-\infty}^{\infty} dz\,  \exp \left[ -\frac{z^2}{2H^2} - \frac{\St_i}{\beta} \left( \exp\frac{z^2}{2H^2} -1\right)\right]},
\EN
where the factor of $2$ accounts for the upper and low halves of the disk, assumed to be symmetric about the midplane.
The ratio between the fractional rates at which the two dust populations are processed to make chondrules is given by
\EQ
\xi \equiv f_2/f_1, \label{xi_def}
\EN
which is a function of $\St_2/\St_1$, $\St_2/\beta$ and $z_c/H$ (or other permutations). Note that, as we have arbitrarily set $\St_1 < \St_2$,
we expect $\xi <1$: the dust grain population $i=2$, made of larger, more settled grains, would have preferentially settled out 
of the chondrule forming region.

Considerations of how much matrix was thermally
processed suggest that only a small fraction of ambient dust was ever processed into chondrules \citep{2015Icar..245...32H}.
As long as that was indeed the case, the ratio of the rates at which the dust populations were melted to make chondrules was also
the ratio of the mass fractions of the populations which were melted into chondrules. Using that assumption we can explore the consequences of $\xi$. 
We assume that population $2$ contained a fraction
$f$ of the total dust mass, and population $1$ contained the rest (i.e.~a fraction $1-f$). We also assume that population $i$ contained
a mass fraction $g_i$ of a given element or isotope of interest. Then the overall mass fraction of that element was simply
\EQ
g_1 (1-f) + g_2 f,
\EN
while the chondrule mass fraction of that element would have been
\EQ
\frac{g_1 (1-f) + \xi g_2 f}{(1-f)+\xi f},
\EN
and the enrichment factor EF of that element compared to bulk ambient dust (which is not the same as future chondrite bulk) would
have been
\EQ
\text{EF} = \left[g_1 (1-f) + g_2 f\right]^{-1}\frac{g_1 (1-f) + \xi g_2 f}{(1-f)+\xi f}.
\EN
Note that depletion is represented by EF$<1$.
In the limiting case of $g_2=0$ we find that the enrichment factor would have been
\EQ
\text{EF}= \frac{1}{(1-f)+\xi f} > 1,
\EN
which tends to $\xi^{-1}$ as $f$ tends to $1$. On the other hand, in the limiting case of $g_1=0$ we have depletion, with
\EQ
\text{EF} = \frac{ \xi }{(1-f)+\xi f}<1,
\EN
which tends to $\xi$ as $f$ tends to $0$.
If all of our element of interest was found in dust grains that were preferentially melted
to make chondrules/preferentially preserved as matrix, then a significant enrichment/depletion of that
element in chondrules with respect to bulk was possible.

In Figures~(\ref{Scale_rat_large_fig}) and (\ref{Scale_rat_small_fig}) we plot the ratio $\xi$ for a variety of parameters.
The contours show the line $\xi=0.5$, at which value twice the fraction of the smaller (lower $\St$) of the two dust agglomerations populations would
have been processed
into chondrules compared to the larger dust agglomeration population. As sketched above, even in the limit of extreme dust elemental or isotopic inhomogeneities,
$\xi < 0.5$ is required for order-unity enrichment or depletion of chondrules with respect to the overall bulk abundances.
As such, $\xi=0.5$ can be taken as an upper limit for values of $\xi$ which could explain compositional variations between chondrules and matrix.
Disk surface layers are expected
to be strongly turbulent, and must have been for turbulent dissipation to be the chondrule melting mechanism.
Accordingly, we can use
\EQ
\beta \sim \alpha \gtrsim 10^{-3}
\EN
as a generous lower bound on $\beta$. As calculated in \Sec{Sec_DVD}, chondrule midplane Stokes numbers were approximately $\St_0 \simeq 3 \times 10^{-4}$,
so we can estimate that porous chondrule precursors had $\St_0 < 10^{-4}$ (i.e. volume filling fractions below $20\%$). It follows that we expect $\St_0/\beta < 0.1$. Even for extreme
values of $\St_2/\St_1$ this forces the constraint that $z_c > H$, as can be seen in \Fig{Scale_rat_large_fig}.

\begin{figure}\begin{center}
\includegraphics[width=\columnwidth]{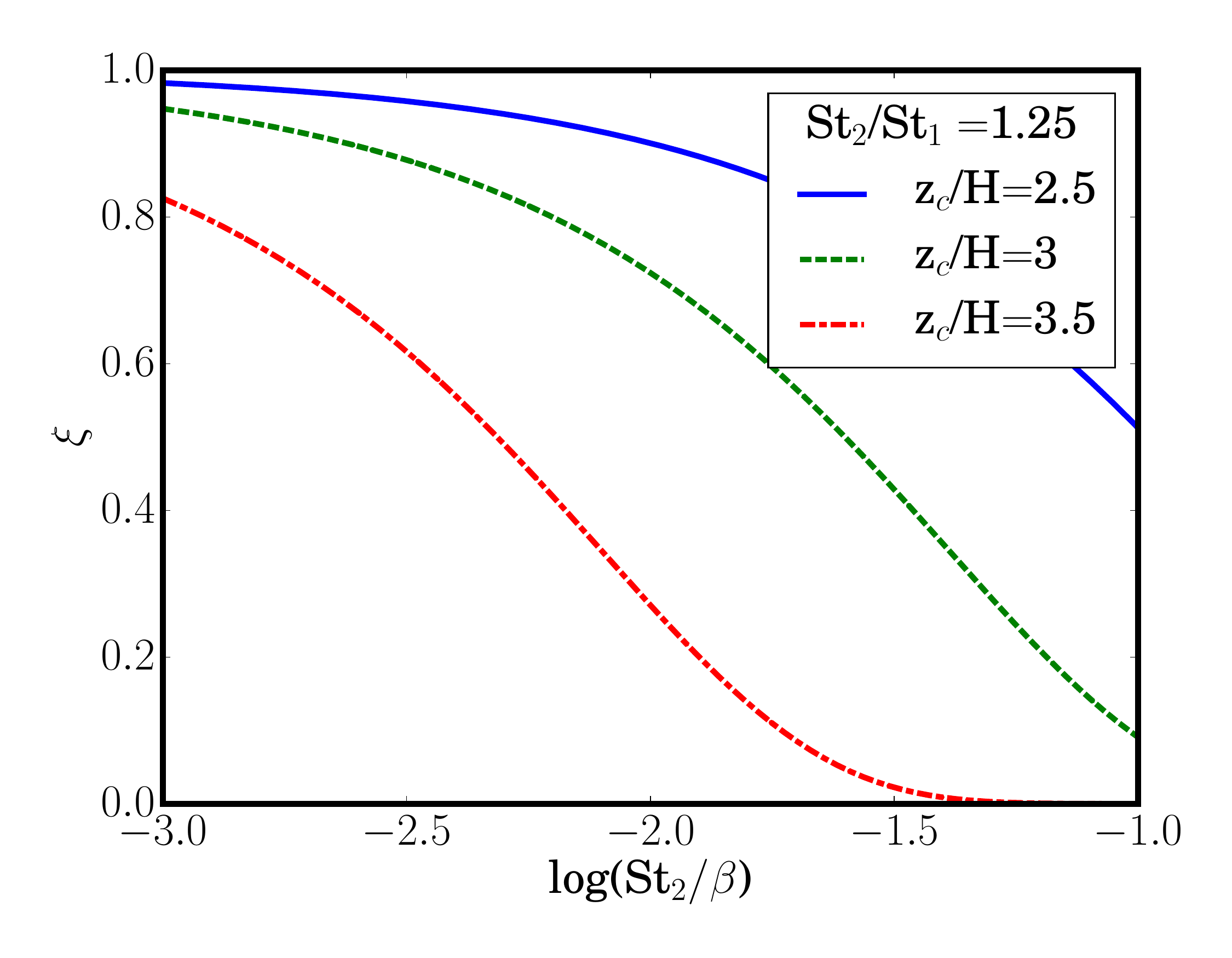}
\end{center}\caption{
Chondrule processing ratio $\xi$ (Equation~\ref{xi_def}) as a function of $\St_2/\beta$ for $\St_2/\St_1 = 1.25$ and three
different values of $z_c/H$. Blue/solid: $z_c/H=2.5$,  green/dashed: $z_c/H=3$, red/dash-dotted:  Blue/solid: $z_c/H=3.5$.
\label{beta_fig} }
\end{figure}

Turbulently stirred dust collides at speeds \citep{1980A&A....85..316V}
\EQ
v_c \sim \sqrt{\beta \St} c_s. \label{v_col}
\EN
For chondrule precursors at the midplane, with $St_0=10^{-4}$, and midplane $\beta=10^{-4}$ in an MMSN at $R=2.5$\,AU with
$c_s \simeq 8 \times 10^4$\,cm\,s$^{-1}$, \Eq{v_col} implies $v_c \sim 8$\,cm\,s$^{-1}$. This lies well between the bouncing
and fragmentation barriers \citep{2010A&A...513A..57Z}, so we expect chondrule precursors to mostly bounce upon collision, and
to be at an equilibrium between infrequent low velocity sticking events and infrequent high velocity fragmentation events pulled from a nearly Maxwellian
turbulent collisional velocity distribution.

\begin{figure}\begin{center}
\includegraphics[width=\columnwidth]{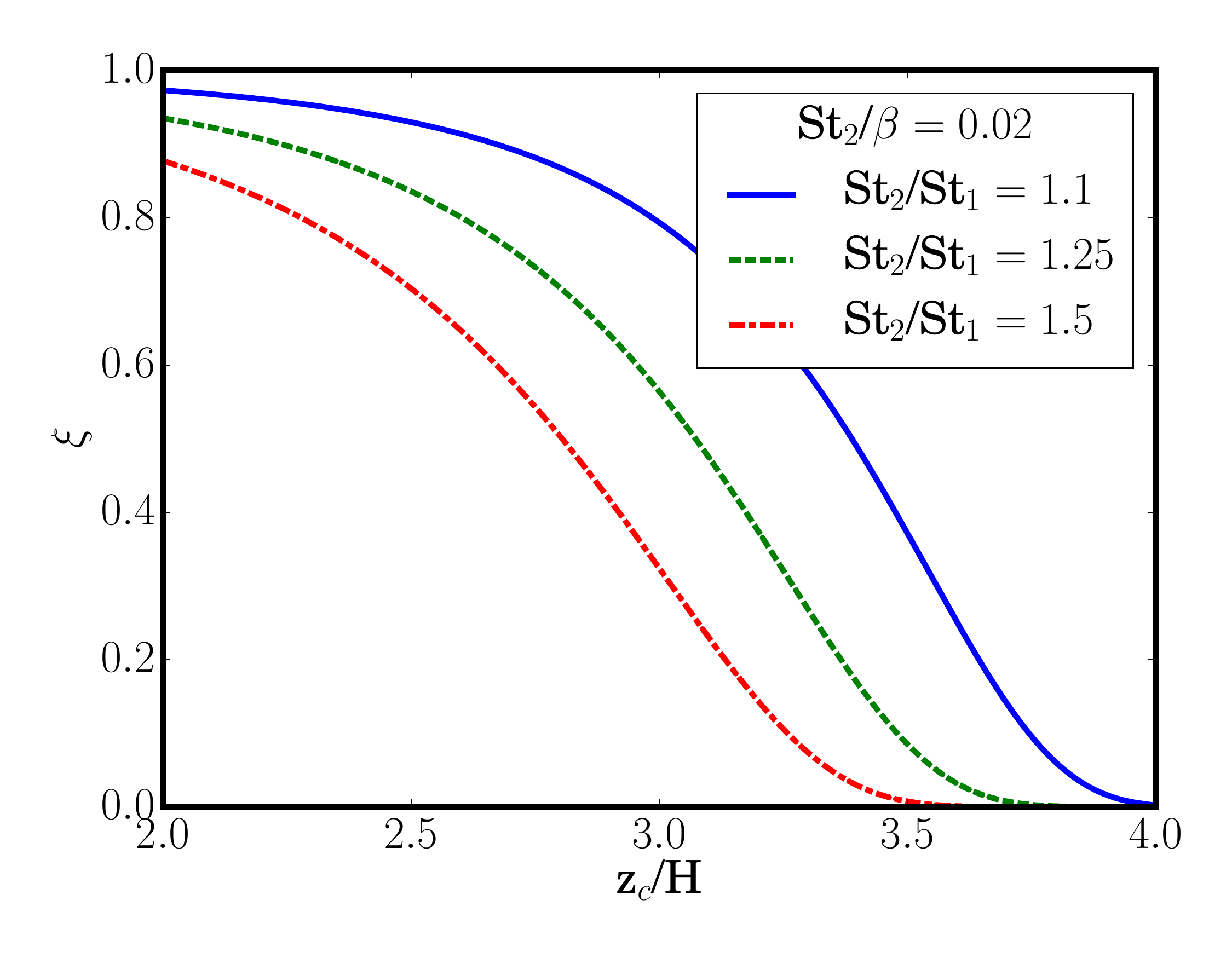}
\end{center}\caption{
Chondrule processing ratio $\xi$ (Equation~\ref{xi_def}) as a function of $z_c/H$ for $\St_2/\beta=0.02$ and three
different Stokes ratios. Blue/solid: $\St_2/\St_1=1.1$,  green/dashed: $\St_2/\St_1=1.25$, red/dash-dotted:  Blue/solid: $\St_2/\St_1=1.5$.
\label{scan_fig} }
\end{figure}

In Paper I we noted that the fragmentation rate is controlled by the tail of the turbulent collisional Maxwellian distribution,
and hence is an exquisitely sensitive function of the collisional velocity scale.
As a consequence, the equilibrium dust size is relatively insensitive to the stickiness of the participants. If that is indeed the controlling factor for
the dust size, then we expect $\St_2 \sim \St_1$.  In \Fig{Scale_rat_small_fig} we can see values of $\xi<0.5$ can indeed be
achieved for $\St_2 / \St_1 <1.5$ and $\St_0/\beta<0.1$, but only for $z_c>2H$, and even that altitude
is generally insufficient as can be seen in \Fig{beta_fig}. Further, at those altitudes, small changes in $z_c/H$ can have
large implications for $\xi$, as shown in \Fig{scan_fig}.

\section{Discussion and Conclusions}

The partitioning of refractory elements and their isotopes between chondrules and matrix in chondritic meteorites implies that
there were elementally and isotopically distinct dust populations, some of which were preferentially melted to make chondrules.
This requires not only that dust collisional coagulation cannot have treated all potential collisional partners the same, but also
requires that there have existed a mechanism which sorted dust grains into chondrule-precursors and matrix-precursors.
If dust collisional coagulation treated potential collisional partners differently, then dust grains with different
compositions would be expected to have grown to (slightly) different sizes and hence aerodynamics.
We discuss this part of the problem in a companion paper \citep{Fe-Col}.

Such a difference in aerodynamics would have allowed the chondrule-or-matrix fate of a dust grain to have been
determined through aerodynamical sorting: different dust agglomerations would have been preferentially found
in differing regions of the disk. In that case, chondrule formation that was spatially localized in a fashion correlated
with the spatial dust sorting would explain how refractory
elements were partitioned between matrix and chondrules.
If, as seems likely,  chemical and thermal processing cannot explain the tungsten isotope partitioning seen by \cite{2015LPI....46.2262B},
then the alternative to
dust sorting are large scale solar nebula composition gradients.
However, \cite{2015MNRAS.452.4054G} showed that matrix and chondrules had to have been in some fashion co-genetic, which means
tapping into a large scale gradient would require extreme fine tuning.

This is particularly pressing for Allende because that meteorite is
tungsten-isotopic-complementary with the inner solar system, while its individual components vary strongly \citep{2016LPI....47.1453B}.
Given the large chondrite to chondrite matrix volume fractions, it is difficult to imagine why Allende's
chondrules and Allende's matrix would have originated
from separate, unmixed dust reservoirs within the solar nebula, and yet still have been subsequently mixed in exactly the correct ratio to produce the observed uniform
inner solar system bulk compositions.

While tungsten is highly refractory, some of the compositional variation between matrix and chondrules is undoubtably due to
partitioning of more volatile elements during chondrule melting by evaporation \citep[for a selection of matrix/chondrule compositional differences, see][]{2016GeCoA.172..322E}.
Further, because chondrules and matrix likely had differing aerodynamics, they were likely consumed by the parent body formation process
at differing rates \citep{2015Icar..245...32H}. This means that even if ambient proto-matrix and chondrules were perfectly correlated in the solar
nebula, they would be imperfectly correlated, especially in the more volatile elements, once combined into a parent body.
Because isotopic signatures in general, and highly refractory tungsten isotopes specifically are 
relatively unaffected by thermal or chemical processes, they offer an excellent complement to the partitioning of more volatile elements
in learning how correlated the matrix and chondrules actually were, and especially how that correlation came to be or was maintained.

We have examined the conditions required for dust spatial sorting and chondrule forming localization to generate significant partitioning.
We found that partitioning is possible, but requires strong sorting comparable to settling at high, $z>2H$, altitude, and more likely regions
of parameter space require $z \gtrsim 3H$. This argues strongly
against vertically stochastic chondrule forming mechanisms such as planetesimal bow shocks
\citep{1998M&PS...33...97H,2012ApJ...752...27M},  but fits well with a turbulent dissipation scenario
for chondrule formation. FUV and X-ray non-thermal ionization, which cannot penetrate deeply,
make protoplanetary disk surface layers more magnetically active
than the cold, neutral midplane \citep{1996ApJ...457..355G}, and hence good candidate chondrule formation regions. Because the high altitude chondrule forming regions
would contain only a small fraction of the overall disk mass, they are also promising in that they would not be expected to process too much disk material,
and could explain why chondrules are as small as they are \citep{2015Icar..245...32H}.  Other locations, such as radially localized
zonal flows, could have
also achieved the required sorting, but would have had to have been extremely strong, and would have had to have been
temporally correlated with chondrule formation for the sorting
to have led to chondrule-matrix partitioning.

\section*{Acknowledgements}
This research was inspired and greatly helped by conversations with Denton S. Ebel.
Insightful comments by the referee greatly refined and clarified the arguments.
The research leading to these results was funded by NASA OSS grant NNX14AJ56G.

\bibliography{Rel_Set}

\end{document}